\documentclass[a4paper,twocolumn,showpacs,preprintnumbers,prl,tightenlines,%
  superscriptaddress,footinbib]{revtex4}
\usepackage[dvips]{graphicx}
\usepackage{dcolumn}
\usepackage{bm}
\usepackage{amsmath}

\usepackage{amsfonts}
\usepackage{color}
\providecommand{\U}[1]{\protect\rule{.1in}{.1in}}
\providecommand{\U}[1]{\protect\rule{.1in}{.1in}} \usepackage{epsfig}

\newcommand{\be}{\begin{equation}}
\newcommand{\ee}{\end{equation}}
\newcommand{\bea}{\begin{eqnarray}}
\newcommand{\eea}{\end{eqnarray}}
\newcommand{\fluck}{\langle k^2\rangle}
\newcommand{\avk}{\langle k\rangle}
\newcommand{\av}[1]{\langle {#1} \rangle}

\begin{document}

\title{Thresholds for epidemic spreading in networks}

\author{Claudio Castellano} 
\affiliation{Istituto dei Sistemi Complessi (CNR-ISC), UOS Sapienza
and Dip. di Fisica, ``Sapienza''
  Universit\`a di Roma, P.le A. Moro 2, I-00185 Roma, Italy}
\author{Romualdo Pastor-Satorras} 
\affiliation{Departament de F\'\i sica i Enginyeria Nuclear,
  Universitat Polit\`ecnica de Catalunya, Campus Nord B4, 08034
  Barcelona, Spain}

\begin{abstract} 
We study the threshold of epidemic models in quenched 
networks with degree distribution given by a power-law.
For the susceptible-infected-susceptible model
the activity threshold $\lambda_c$ vanishes in
the large size limit on any network whose maximum degree
$k_{max}$ diverges with the system size,
 at odds with heterogeneous
mean-field (HMF) theory.
The vanishing of the threshold has nothing to do with the scale-free
nature of the network but stems instead from the
largest hub in the system being active for any spreading rate
$\lambda>1/\sqrt{k_{max}}$ and playing the role of a self-sustained
source that spreads the infection to the rest of the
system.
The susceptible-infected-removed model displays
instead agreement with HMF theory and a finite threshold for scale-rich
networks. We conjecture that on quenched scale-rich networks 
the threshold of generic epidemic models is vanishing or finite depending
on the presence or absence of a steady state.
\end{abstract}
\maketitle

The heterogeneous pattern of a network can have dramatic effects on
the behavior of dynamical processes running on top of it
\cite{barratbook}, in particular when the distribution of the number
$k$ of contacts (the degree of an element or vertex) exhibits long
tails, as expressed by a power-law degree probability with the
asymptotic form $P(k) \sim k^{-\gamma}$ \cite{barabasi02}. An example
that has attracted a great interest due to its practical real-world
implications is the modeling of epidemic spreading on contact networks
\cite{keeling05:_networ}.  The simplest of these models is the SIS
model \cite{anderson92}, in which each vertex (individual) can be in
one of two states, either susceptible, or infected.  Susceptibles
become infected by contact with infected individuals, with a rate
proportional to the number of infected contacts times a given
spreading rate $\lambda$. Infected individuals on the other hand
become healthy again with a rate that can be set arbitrarily equal to
unity. The model allows thus individuals to contract the infection
time and again, leading, in the infinite network size limit, to a
sustained infected steady state for values of $\lambda$ larger than an
epidemic threshold $\lambda_c$.  In the 
SIR model \cite{anderson92}, on the other hand, infected individuals
recover (or die) and cannot change further their state.  No steady
state is now allowed, but a threshold still exists above which the
total number of infected individuals, starting from a very small
infected seed, reaches a finite fraction of the network.  The analysis
of these and other models \cite{barratbook}, performed via a
mean-field theory modified to take into account the heterogeneity of
the network substrate \cite{Pastor01, dorogovtsev07:_critic_phenom},
led to the far-reaching conclusion that topological fluctuations, as
measured by the second moment of the degree distribution $\fluck$, can
have profound effects in many types of dynamics
\cite{dorogovtsev07:_critic_phenom,barratbook}. Thus, for example, in
the SIS model, the threshold takes the values, at the mean-field level
$\lambda_c = \avk/\fluck$.
For a long-tailed degree distribution with power law form, the second
moment diverges for $\gamma\leq 3$, and one obtains the remarkable
result of a vanishing epidemic threshold in the thermodynamic
limit. These results have led to the widespread belief 
in the distinction between \emph{scale-free}
networks with $\gamma\leq 3$, where topology is highly relevant, 
and
\emph{scale-rich} networks with $\gamma>3$, where dynamical processes
exhibit an essentially homogeneous mean-field behavior.

In this Letter, building on some results 
previously reported,
we present evidence that this
belief is not correct for the SIS model on quenched networks
(i.e. networks whose adjacency matrix is fixed in time)
and that the scale-free nature
of the contact pattern has no crucial effect on the value of the
epidemic threshold.  We investigate the physical origin of
this result, its validity for generic network structures and its
consequences.
On the other hand we show that for the SIR model the picture is
different, a zero threshold occurring only in scale-free
quenched networks.

While heterogeneous mean-field (HMF) theory  is exact on annealed
networks 
(i.e. networks 
whose adjacency matrix in fixed only in average
\cite{dorogovtsev07:_critic_phenom}),
results beyond HMF theory for the SIS process on quenched networks (QN)
have appeared in different contexts and with various levels of rigor.
Already in 2003, Wang et al.~\cite{Wang03} argued that the
epidemic threshold on an {\em arbitrary} undirected graph is set by the
largest eigenvalue $\Lambda_N$ of the adjacency matrix, 
\begin{equation}
  \lambda_c = 
  \Lambda_N^{-1},
  \label{lambda_c}
\end{equation}
see also
\cite{0295-5075-89-3-38009,Prakash:1257394}.
The relevance of Eq.~(\ref{lambda_c})
becomes evident when it is complemented with the results of Chung et
al.~\cite{Chung03}, who calculated the largest eigenvalue of the
adjacency matrix for a class of finite graphs with
degrees distributed according to a power-law, obtaining
\begin{equation}
  \Lambda_N = 
  \left \{ 
    \begin{array}{lr}
      c_1 \sqrt{k_{c}} &~~~~~~~ \sqrt{k_{c}} > \frac{\fluck}{\avk}
      \ln^2(N) \\ 
      c_2 \frac{\fluck}{\avk} & ~~~~~~~\frac{\fluck}{\avk} >
      \sqrt{k_{c}} \ln(N) 
    \end{array}
  \right. ,
  \label{Lambda_N}
\end{equation}
where $N$ is the network size, $k_{c}$ is network cut-off or degree of the most
connected node (averaged over many network realizations
\cite{Restrepo07}), and $c_i$ are constants of order 1. 
The cut-off $k_c$ is a growing function of the network size for uncorrelated
scale-free networks, taking the value $k_c \sim N^{1/2}$ for
$\gamma\leq3$ and $k_c \sim N^{1/(\gamma-1)}$ for $\gamma > 3$
\cite{mariancutofss}.  For $\gamma>3$ the ratio of the moments is
finite and it is clear that the largest eigenvalue is governed by
$k_{c}$.  Noticeably this remains true also for $5/2 <\gamma < 3$,
since in that range $\fluck/\avk \sim k_{c}^{3-\gamma} \ll
\sqrt{k_{c}}$.  Only for $2 < \gamma < 5/2$ the largest eigenvalue is
set by the moments of the degree distribution.  Combining
Eqs.~(\ref{lambda_c}) and~(\ref{Lambda_N}), the behavior of the
threshold for the SIS model in a power-law distributed
network is, for sufficiently large size,
\begin{equation}
  \lambda_c
  \simeq \left \{
    \begin{array}{lr}
      1/\sqrt{k_{c}} & ~~~~~~~ \  \
      \gamma > 5/2 \\
      \frac{\avk}{\fluck} & ~~~~~~~ 2< \gamma < 5/2
    \end{array}
  \right. ,
  \label{together}
\end{equation}
see also \cite{Ganesh05}. 
Since $k_c$ grows as a function of $N$ for any $\gamma$, the
consequence of Eq.~\eqref{together} is remarkable: {\em In any
uncorrelated quenched random network with power-law distributed
connectivities, the epidemic threshold for SIS goes to zero as the system
  size goes to infinity}. This has nothing to do with the scale-free
nature of the degree distribution: It is always true as long as the
cut-off $k_c$ diverges.
Remarkably the threshold goes to zero also for Erd\"os-R\'enyi
graphs (although logarithmically slow),
for which a formula similar to Eq.~(\ref{Lambda_N})
exists~\cite{Krivelevich03}.  
Different approaches \cite{Ganesh05,Chatterjee09,durret10:_some},
have also pointed out that in the thermodynamic limit, the
system is active for any $\lambda>0$. 
These results, however, have gone largely unnoticed within the
statistical physics community.

A first issue raised by Eq.~\eqref{together} concerns the fact that,
as any critical point, the epidemic threshold is well defined only in
the thermodynamic limit.
In a finite system, the dynamics is always doomed to fall into the healthy,
absorbing state, even far above the threshold, due to stochastic
fluctuations.  The threshold for a finite network of size $N$
must therefore be intended as the value separating the regime
$\lambda<\lambda_c$ for which the epidemics decays exponentially fast
(so that the expected survival time is of the order $\tau \sim
\ln(N)$) from the regime $\lambda>\lambda_c$ where the survival time
grows exponentially with $N$ to some power, $\tau \sim e^{N^\alpha}$,
with $\alpha>0$.

\begin{figure}
  \begin{center}
    \includegraphics[width=7.5cm,angle=0]{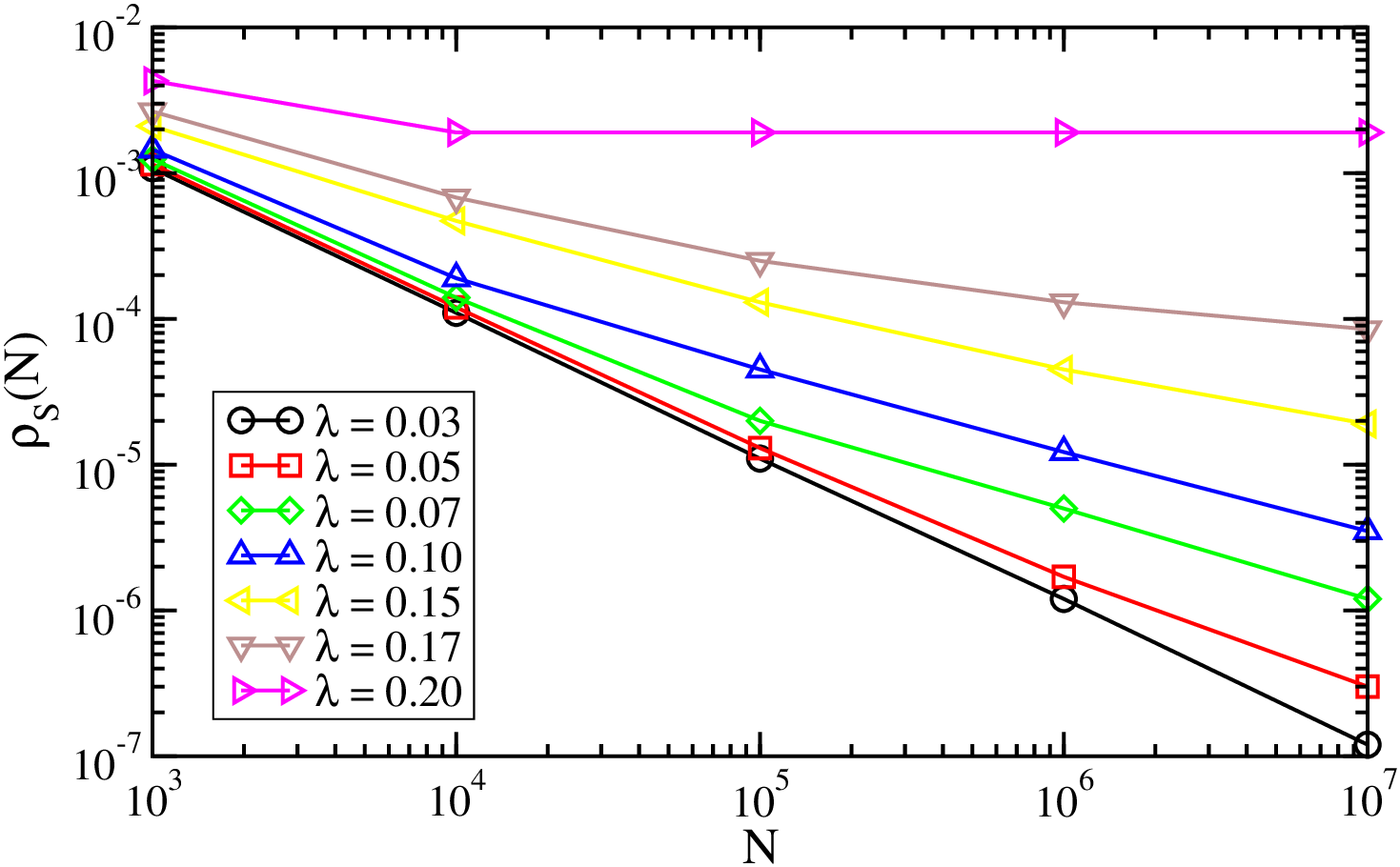}
    \caption{Density of active sites for long times (restricted to
      surviving runs) in the SIS model on QN as a
      function of system size $N$, for $\gamma=4.5$
      and different
      values of the parameter $\lambda$. Notice that the straight line
      for $\lambda=0.03$ is due to the fact that no density smaller
      than $1/N$ can occur.}
    \label{FSS}
	\end{center}
\end{figure}

To investigate the validity of these results we have performed
numerical simulations of the SIS model on quenched scale-rich networks with
$\gamma=4.5$ and minimum degree $k_{min}=3$, built using the
uncorrelated configuration model~\cite{Catanzaro05}.  In order to
compare results with the predictions in Eq.~\eqref{together} one must
take into account that the actual maximum degree $k_{max}$ in each
network realization is a random variable, with average value
$\av{k_{max}} = k_c$. In particular, in the case $\gamma>3$,
one can see \cite{Motter07} that both the mean and
the standard deviation of $k_{max}$ scale as $k_c \sim
N^{1/(\gamma-1)}$, implying that $k_{max}$ always shows large fluctuations
for different realizations of the degree sequence. Therefore, we
first consider networks in which $k_{max}$ has a fixed value, equal to
the mean $k_{c}$ numerically estimated for the chosen system size $N$.
In Fig.~\ref{FSS} we plot the density $\rho_s$, calculated only for
surviving runs, as a function of $N$ for different values of
$\lambda$ \cite{marro99}.
Should the transition occur at a fixed value of $\lambda$, $\rho_s$
would go to a constant for $\lambda>\lambda_c$, decay exponentially
for $\lambda<\lambda_c$ and as a power-law exactly at the transition.
A completely different behavior is observed:
for all values of $\lambda$, the curves are bent upward, indicating
that the system is active for any $\lambda$.  This
excludes the presence of a finite threshold for diverging $N$.
While Eq.~(\ref{lambda_c}) holds for SIS on any graph,
Eq.~\eqref{Lambda_N} was instead obtained for a specific network model
(intrinsically correlated for $\gamma<3$ and uncorrelated for
$\gamma>3$ \cite{mariancutofss}).
For generic topologies, it is simple to show~\cite{Restrepo07}
that $\sqrt{k_{c}}$ is a lower bound for the
largest eigenvalue of the adjacency matrix.  This allows to conclude that,
unless the degree distribution is strictly bounded from above,
the threshold for SIS on any graph vanishes in the
thermodynamic limit.

\begin{figure}
  \begin{center}
    \includegraphics[height=4.5cm,angle=0]{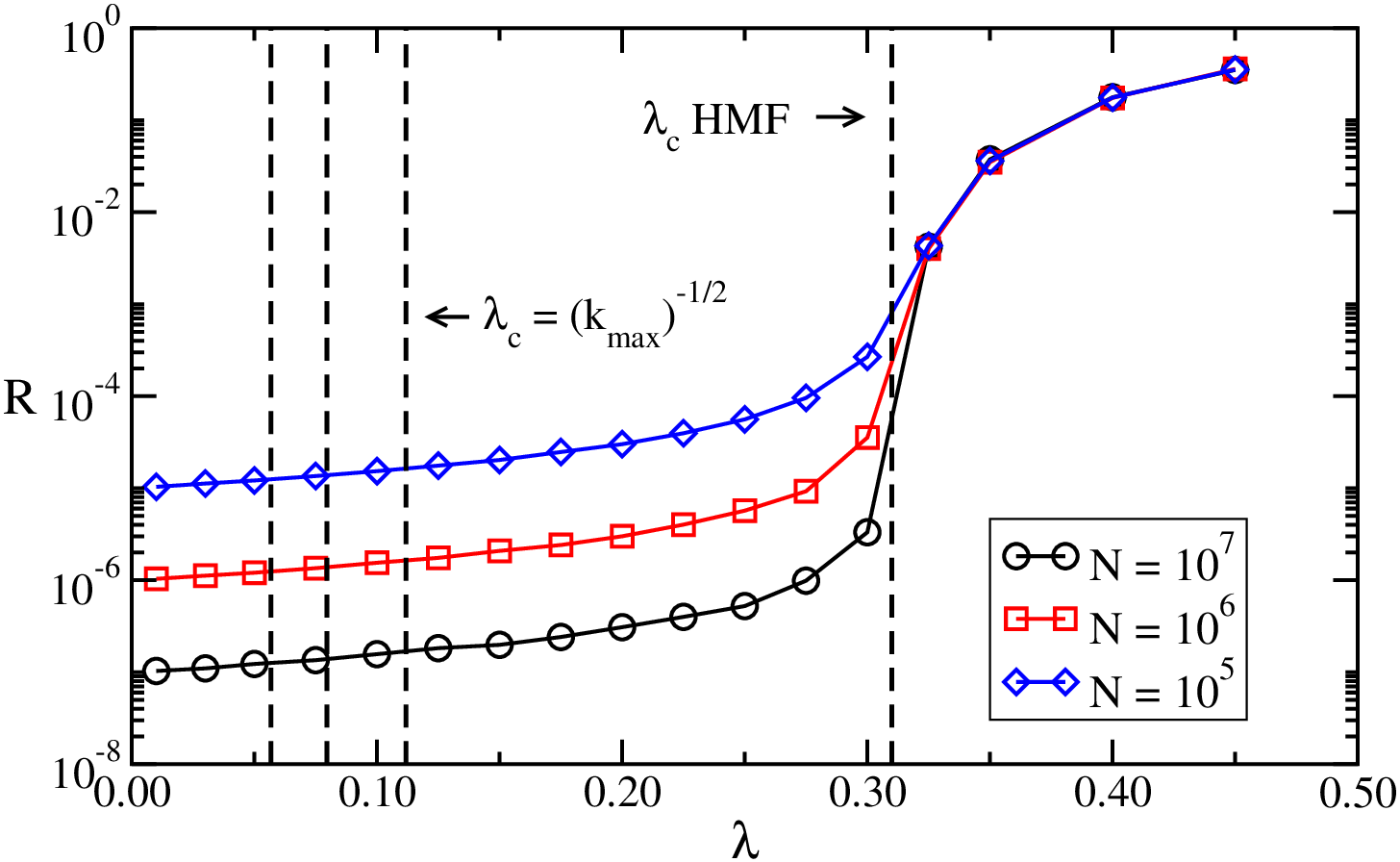}
    \caption{Total number $R$ of infected individuals in the SIR model
      on QN of different size $N$ as a function of the
      spreading rate $\lambda$.  Networks have $\gamma=4.5$.}
    \label{SIR}
  \end{center}
\end{figure}

How generic are these results?
Prakash et al.~\cite{Prakash:1257394} have recently argued that
Eq.~\eqref{lambda_c} is valid for \textit{all}
epidemic processes, regardless of their particular microscopic details.
To check this claim, we consider the SIR model.
At the HMF level, the threshold takes
the value $\lambda_c^{SIR} = \av{k}/[\fluck - \avk]$ \cite{Cohen00,newman02}
and is therefore finite for scale-rich networks with $\gamma>3$. From the
analysis in Eq.~\eqref{together}, on the other hand, it should be
vanishingly small in the large network limit, according to
Ref.~\cite{Prakash:1257394}. We have checked this possibility by performing
numerical simulations of the SIR model on networks with $\gamma=4.5$ and
different values of $N$, with fixed $k_{max} = \av{k_{max}}$.
In this case, the HMF estimated threshold takes the value
$\lambda_c^{SIR} \simeq 0.31$, independent of the network size, while the
predictions from Eq.~\eqref{together} are $\lambda_c^{SIR}\simeq
0.0567$, $0.0796$, and $0.1118$ for the different network sizes considered.
In Fig.~\ref{SIR} we report the final density of infected
individuals $R$ as a function of the spreading rate $\lambda$,
starting from a single randomly chosen infected node.
The prediction of HMF theory seems in this case to be much more accurate
than Eq.~\eqref{together}, contrary to the generic claim made in
Ref.~\cite{Prakash:1257394}: The threshold remains finite in the
large $N$ limit.

\begin{figure}
  \begin{center}
    \includegraphics[height=4.5cm,angle=0]{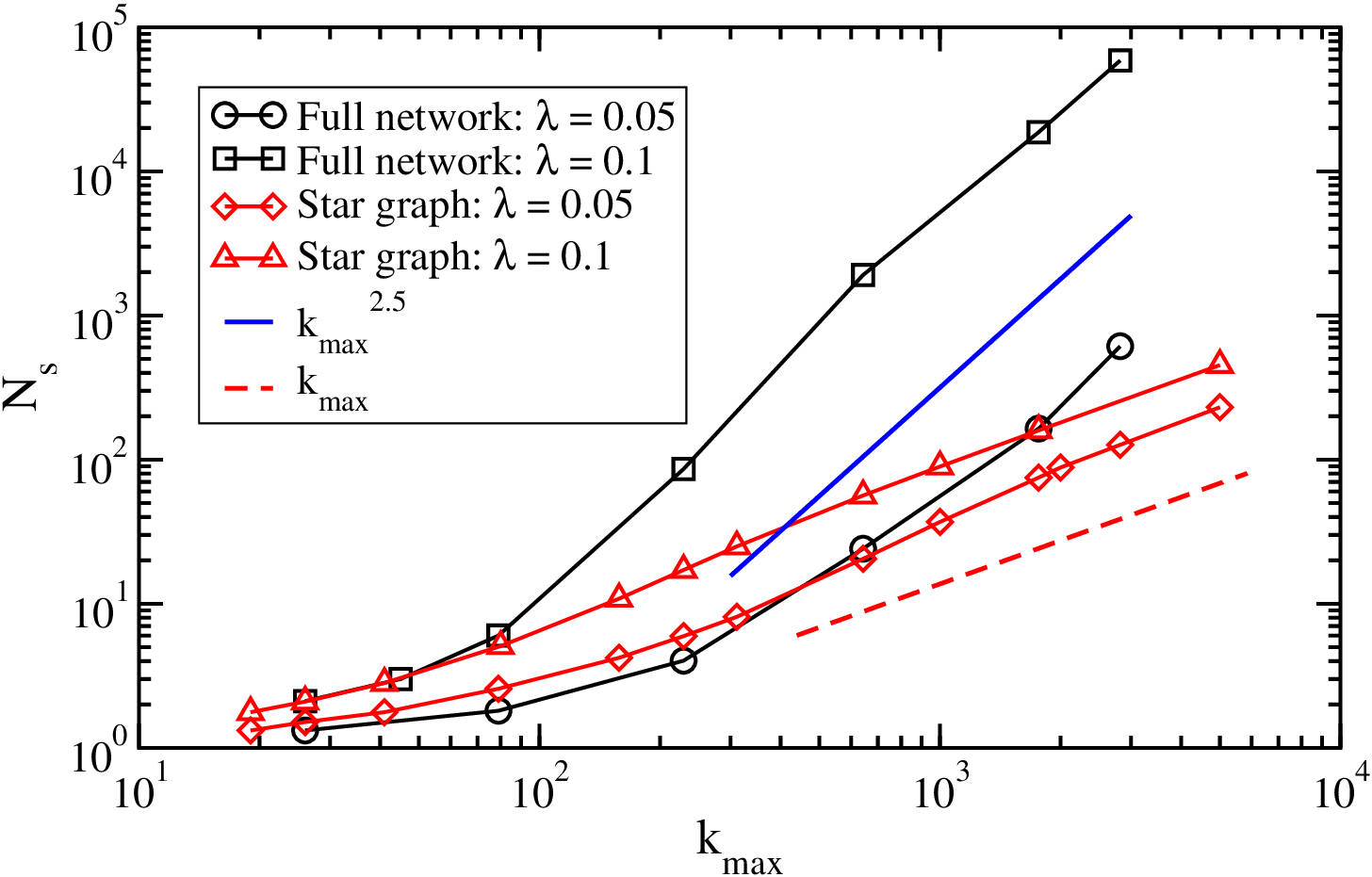}
    \caption{Number $N_s$ of active nodes in surviving runs for
      values of $\lambda$ smaller than the threshold predicted by HMF
      ($\lambda_c(HMF)=0.138\ldots$) as a function of $k_{max}$,
      compared with the same quantity for star
      graphs. Networks have $\gamma=3.5$.}
    \label{Figstar}
\end{center}
\end{figure}

To understand 
the different behavior of the
two models we look at 
the origin of the
incorrect HMF prediction for SIS in QN.
From a mathematical point of view, the HMF approach is
equivalent~\cite{dorogovtsev07:_critic_phenom} to replacing the
QN with given adjacency matrix $a_{ij}$ by an annealed
network with an averaged adjacency matrix, $\bar{a}_{ij}$
\cite{dorogovtsev07:_critic_phenom,Boguna08}.  In the uncorrelated
case this matrix reduces to
$\bar{a}_{ij}= k_i k_j/[N \avk]$, which
has a unique nonzero eigenvalue $\bar{\Lambda}_N = \fluck/\avk$.
Hence the annealed network approximation destroys the detailed 
structure of the eigenvalue spectrum of QN
and preserves the correct largest eigenvalue only for $\gamma<5/2$.
This basic feature, and not (as suggested in Ref.~\cite{durret10:_some})
the disregard of dynamical correlations, is at root of the inaccuracy of
the HMF approach. 
A more physical  insight comes from the
analysis  of a star graph with one center connected to $k_{max}$ leaves
of degree $1$.
In this case the largest eigenvalue of the adjacency matrix
is $\Lambda_N = \sqrt{k_{max}}$
which implies $\lambda_c = 1/\sqrt{k_{max}}$.  
The same result can be easily recovered 
by writing the rate equations for the probability $\rho_{max}$
($\rho_1$) for the center (leaves) to be active, namely
$\dot{\rho}_{max}  =  -\rho_{max} + (1-\rho_{max}) \rho_1 \lambda k_{max}$
and 
$\dot{\rho_1} = -\rho_1 + (1-\rho_1) \rho_{max} \lambda$.
Imposing the steady state condition one finds 
\begin{eqnarray}
  \rho_{max} =
  \frac{\lambda^2 k_{max}-1}{(1+\lambda k_{max}) \lambda} \qquad
  \rho_1 = \frac{\lambda^2 k_{max}-1}{(1+\lambda)k_{max} \lambda},
  \label{eq:4}
\end{eqnarray}
and hence the threshold condition above.  The message of
Eq.~\eqref{eq:4} for a generic quenched random graph is strong:
Independently from all the rest of the system, for $\lambda
>1/\sqrt{k_{max}}$ the subgraph composed by the node with degree
$k_{max}$ and its neighbors is in the active state.  This core of
activity provides a self-sustained source of infection that, since in
the full graph the neighbors of the hub are not leaves, can transfer
the activity to their other neighbors and spread in this way the
epidemics to a finite fraction of vertices.  This is confirmed by
Fig.~\ref{Figstar}, showing the number of actives nodes in
surviving runs, $N_s$, on a full network with $\gamma=3.5$ and
on a star graph with the same $k_{max}$\footnote{Analogous results are
obtained for different values of $\gamma$ (data not shown)}.
For $k_{max}<1/\lambda^2$ the values of $N_s$ in the full network and
in the star graph are comparable: both systems are subcritical and the
subgraph centered around the node with degree $k_{max}$ is where
activity lingers before disappearing.  For $k_{max}>1/\lambda^2$ the
star graph becomes active and $N_s$ becomes asymptotically
proportional to $k_{max}$.  In the full network instead, the
asymptotic behavior is $N_s \sim k_{max}^{\gamma-1} \sim N$,
indicating that the active state is endemic: the hub spreads the
activity to a finite fraction of the whole system.  Reaching the
the fully endemic state requires larger systems for small $\lambda$,
but nothing changes qualitatively for any $\lambda>0$.

Understanding the behavior of SIS allows also to unravel why things
go differently for SIR. In the former case, the possibility for 
hubs to be reinfected multiple times, which allows the presence of a
steady state, boosts their impact on the dynamics.
In the case of SIR, on the other hand, high-degree vertices can only
be infected once and this strongly limits their role in the dynamics.
Based on this observation, it is natural to conjecture that epidemic
models allowing a steady state, such as SIS, will lead to a null
threshold in any infinite QN, while all models without
a steady-state will conform with HMF theory, with a finite threshold
on scale-rich topologies.
\begin{figure}
\begin{center}
\includegraphics[width=7.5cm,angle=0]{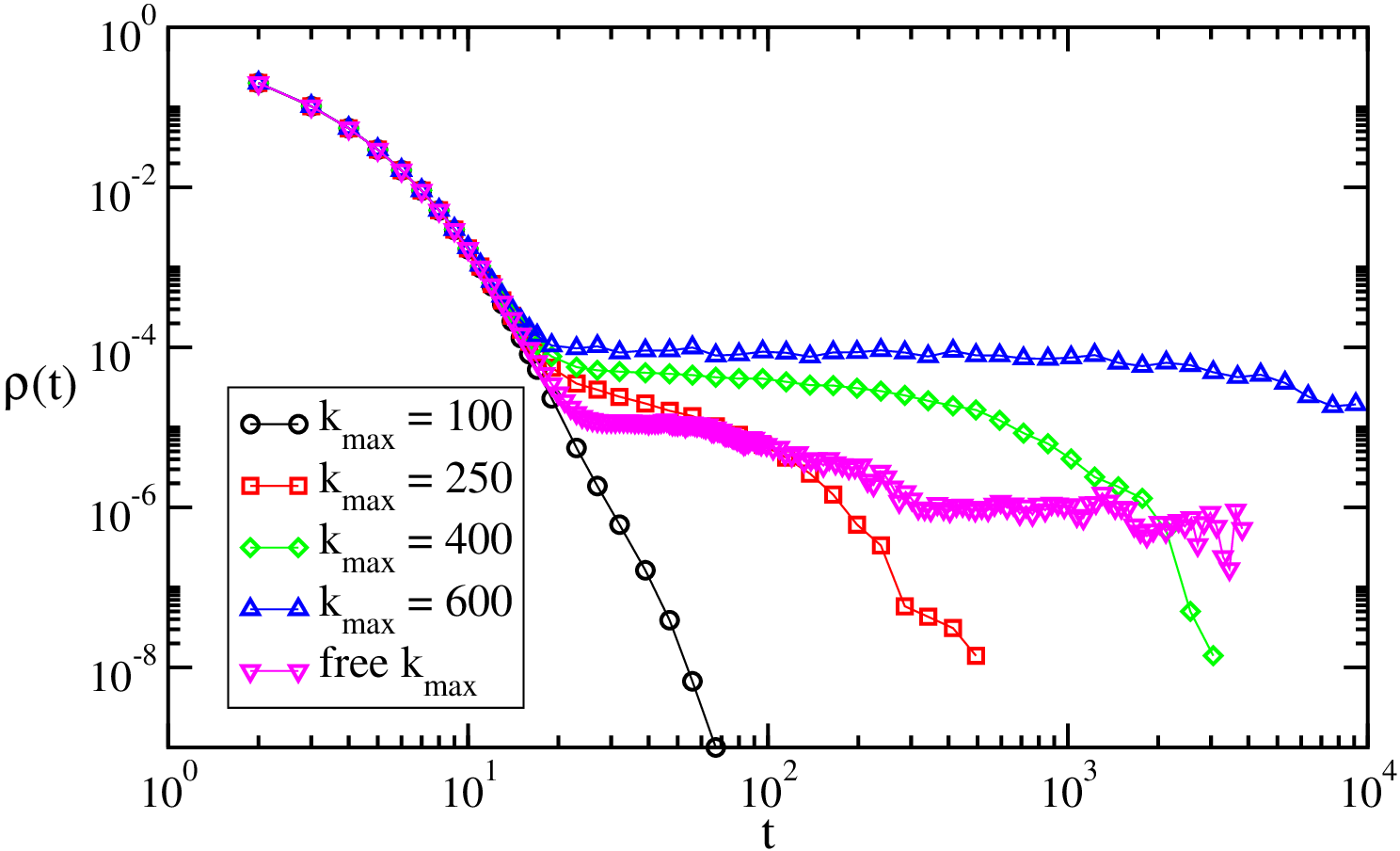}
\caption{Decay of the activity density for the SIS model in
networks with $\gamma=4.5$, $N=10^6$, 
$\lambda = 0.1$ and changing $k_{max}$.}
\label{variouskmax}
\end{center}
\end{figure}

The strong effect of the hub in the dynamics raises further issues
on the SIS model.
While fixing the value of $k_{max}$ to its ensemble average leads to
results consistent with the presence of a non-zero threshold in finite
systems, as implied by Eq.~\eqref{together}, if this constraint
is relaxed, $k_{max}$ has large sample to sample fluctuations leading
to nontrivial consequences.
In Fig.~\ref{variouskmax} we explore the effect of this variability by
comparing simulations performed at fixed $\lambda$ and $N$, and
different values of $k_{max}$. The growth of the activity density
for increasing $k_{max}$ indicates
that the relation between the threshold (or the largest eigenvalue)
and cut-off $k_c$, Eq.~\eqref{together}, can in fact be refined, and
be expressed in terms of the actual maximum degree,
$\lambda_c = 1/\sqrt{k_{max}}$ \cite{Motter07}.  However, the large variations
of $k_{max}$ among different realizations of the network with the same
$\gamma$ and $N$ do not wash away as $N$ diverges and severely
hinder the determination of the threshold in simulations
with unrestricted $k_{max}$. As mentioned before, for $\gamma>3$ the
standard deviation of
$k_{max}$ increases as the average value $\langle k_{max} \rangle \sim
N^{1/(\gamma-1)}$~\cite{Motter07}, and there is always
a large sample to sample variability. 
Hence an unrestricted sampling at fixed $\lambda$ unwarrantedly
averages networks with different thresholds and effective time
scales, some subcritical and some supercritical, making impossible
even to determine the presence of a well-defined steady state.
This fact is exemplified in Fig.~\ref{variouskmax}, where we
plot for comparison the activity density obtained averaging over
networks with a freely varying $k_{max}$.
For $\gamma<3$ the situation depends on
the way the network is generated, and in particular on the way the
upper bound of the degree distribution $M = N^{1/\omega}$ grows
\cite{Boguna08}. If $\omega = 2$ (uncorrelated configuration
model~\cite{Catanzaro05}) or larger, the quantity $\fluck/\avk$
becomes sharply peaked as $N$ grows~\cite{Boguna08}.  If instead
$\omega=1$, as in the normal configuration model~\cite{Molloy95}, the
ratio $\fluck/\avk$ (and hence the threshold) wildly changes from
realization to realization, with relative fluctuations diverging as
$N^{2(3-\gamma)(\gamma-2)/(\gamma-1)}$ \cite{Boguna08}.  Notice that
in the intermediate region $5/2<\gamma<3$, the average value of
$\sqrt{k_{max}}$ is larger than $\fluck/\avk$ but, since fluctuations
of the latter diverge, for some network realizations the actual threshold
$\lambda_c$ is much smaller than the value predicted by
Eq.~(\ref{together}).  We conclude that, unless $\gamma<3$ and $\omega
\ge 2$, no average epidemic threshold can be properly defined from
a numerical point of view for networks with unrestricted $k_{max}$. 

In summary, we have studied how the threshold for models of
epidemic spreading on quenched scale-rich networks behaves as their size grows.
The threshold for SIS model always vanishes
in the thermodynamic limit, due to the role of hubs.
This bears no relationship, at odds with the predictions of HMF theory,
with the divergence of the second moment of the degree distribution,
which is finite.
For the SIR model instead the threshold vanishes only for scale-free
topologies (either quenched or annealed), in agreement with HMF theory.
We conjecture that these different types of behavior are generic for
systems possessing (or not) a steady state.
While the result of a vanishing threshold for SIS is exact
on \emph{quenched} networks, it is however of limited interest
from an epidemiological perspective. The interaction patterns
over which real diseases spread generally vary over short time
scales \cite{butts:revisiting}, and are therefore better described 
by annealed topologies \cite{dorogovtsev07:_critic_phenom}, for
which HMF theory works by definition, and the threshold
is finite for $\gamma>3$.
From a statistical physics point of view, instead, our results
open a promising path towards a better understanding of the scope and
limits of HMF theory as a theoretical tool
to analyze dynamics on heterogeneous networks.

\begin{acknowledgments}
  R.P.-S. acknowledges financial support from the Spanish MEC (FEDER),
  under projects No. FIS2007-66485-C02-01 and FIS2010-21781-C02-01;
  ICREA Academia, funded by the Generalitat de Catalunya; and the
  Junta de Andaluc\'{i}a, under project No. P09-FQM4682.  We thank
  A. Vespignani and M. A. Mu\~noz for discussions.
\end{acknowledgments}



\end{document}